\documentclass[aps,prb,superscriptaddress,shortbibliography,twocolumn]{revtex4-2}

\usepackage{lmodern}

\usepackage{inputenc}
\usepackage{amsmath}
\usepackage{amssymb}
\usepackage{graphicx}
\usepackage{epstopdf}
\usepackage{color}
\usepackage{enumerate}
\usepackage{ulem}

\usepackage{units}
\usepackage{float}
\usepackage{amsmath}
\usepackage{hyperref}
\usepackage{epsfig}
\usepackage{bm}
\usepackage{hyphenat}
\usepackage{makecell}
\usepackage{color,soul}
\usepackage[dvipsnames]{xcolor}

\usepackage{chemformula}

\begin{document}

\title{Saturation of the anomalous Hall effect at high magnetic fields in altermagnetic \ch{RuO2}}

\author{Teresa Tschirner}
\affiliation{Leibniz Institute for Solid State and Materials Research, IFW Dresden, Helmholtzstr. 20, 01069 Dresden, Germany}
\affiliation{W\"urzburg-Dresden Cluster of Excellence ct.qmat, Germany}
\author{Philipp Ke\ss ler}
\affiliation{W\"urzburg-Dresden Cluster of Excellence ct.qmat, Germany}
\affiliation{Physikalisches Institut, Universit\"at W\"urzburg, D-97074 W\"urzburg, Germany}
\author{Ruben Dario Gonzalez Betancourt}
\affiliation{Leibniz Institute for Solid State and Materials Research, IFW Dresden, Helmholtzstr. 20, 01069 Dresden, Germany}
\affiliation{Institute of Physics ASCR, v.v.i., Cukrovarnick\'a 10, 162 53, Prague, Czech Republic}
\author{Tommy Kotte}
\affiliation{Hochfeld-Magnetlabor Dresden (HLD-EMFL), Helmholtz-Zentrum Dresden-Rossendorf, 01328 Dresden, Germany}
\author{Dominik Kriegner}
\affiliation{Institute of Physics ASCR, v.v.i., Cukrovarnick\'a 10, 162 53, Prague, Czech Republic}
\affiliation{Institut f\"ur Festk\"orper- und Materialphysik, Technische Universit\"at Dresden, 01069 Dresden, Germany}
\author{Bernd B\"{u}chner}
\affiliation{Leibniz Institute for Solid State and Materials Research, IFW Dresden, Helmholtzstr. 20, 01069 Dresden, Germany}
\affiliation{W\"urzburg-Dresden Cluster of Excellence ct.qmat, Germany}
\affiliation{Institute of Solid State and Materials Physics, TU Dresden, 01069 Dresden, Germany}
\author{Joseph Dufouleur}
\affiliation{Leibniz Institute for Solid State and Materials Research, IFW Dresden, Helmholtzstr. 20, 01069 Dresden, Germany}
\author{\textcolor{black}{Martin Kamp}}
\affiliation{Physikalisches Institut, Universit\"at W\"urzburg, D-97074 W\"urzburg, Germany; R\"ontgen Center for Complex Material Systems, Universit\"at W\"urzburg, D-97074 W\"urzburg, Germany}
\author{\textcolor{black}{Vedran Jovic}}
\affiliation{Earth Resources and Materials, Institute of Geological and Nuclear Science, Lower Hutt 5010, New Zealand; MacDiarmid Institute for Advanced Materials and Nanotechnology, Wellington 6012, New Zealand;}
\author{Libor Smejkal}
\affiliation{Institut f\"ur Physik, Johannes Gutenberg Universit\"at Mainz, 55128 Mainz, Germany}
\affiliation{Institute of Physics ASCR, v.v.i., Cukrovarnick\'a 10, 162 53, Prague, Czech Republic}
\author{Jairo Sinova}
\affiliation{Institut f\"ur Physik, Johannes Gutenberg Universit\"at Mainz, 55128 Mainz, Germany}
\affiliation{Institute of Physics ASCR, v.v.i., Cukrovarnick\'a 10, 162 53, Prague, Czech Republic}
\author{Ralph Claessen}
\affiliation{W\"urzburg-Dresden Cluster of Excellence ct.qmat, Germany}
\affiliation{Physikalisches Institut, Universit\"at W\"urzburg, D-97074 W\"urzburg, Germany}
\author{Tomas Jungwirth}
\affiliation{Institute of Physics ASCR, v.v.i., Cukrovarnick\'a 10, 162 53, Prague, Czech Republic}
\affiliation{School of Physics and Astronomy, University of Nottingham, NG7 2RD, Nottingham, United Kingdom}
\author{Simon Moser}
\affiliation{W\"urzburg-Dresden Cluster of Excellence ct.qmat, Germany}
\affiliation{Physikalisches Institut, Universit\"at W\"urzburg, D-97074 W\"urzburg, Germany}
\author{Helena Reichlova}
\affiliation{Institute of Physics ASCR, v.v.i., Cukrovarnick\'a 10, 162 53, Prague, Czech Republic}
\affiliation{Institut f\"ur Festk\"orper- und Materialphysik, Technische Universit\"at Dresden, 01069 Dresden, Germany}
\author{Louis Veyrat}
\affiliation{Leibniz Institute for Solid State and Materials Research, IFW Dresden, Helmholtzstr. 20, 01069 Dresden, Germany}
\affiliation{W\"urzburg-Dresden Cluster of Excellence ct.qmat, Germany}
\affiliation{Physikalisches Institut, Universit\"at W\"urzburg, D-97074 W\"urzburg, Germany}

\date{\today}

\begin{abstract}

Observations of the anomalous Hall effect in RuO$_2$ and MnTe have demonstrated unconventional time-reversal symmetry breaking in the electronic structure of a recently identified new class of compensated collinear magnets, dubbed altermagnets. While in MnTe the unconventional anomalous Hall signal accompanied by a vanishing magnetization is observable at remanence, the anomalous Hall effect in RuO$_2$ is excluded by symmetry for the N\'eel vector pointing along the zero-field [001] easy-axis. Guided by a symmetry analysis and ab initio calculations, a field-induced reorientation of the N\'eel vector from the easy-axis  towards the [110] hard-axis was used to demonstrate the anomalous Hall signal in this altermagnet. We confirm the existence of an anomalous Hall effect in our RuO$_2$ thin-film samples whose set of magnetic and magneto-transport characteristics is consistent with the earlier report. By performing our measurements at extreme magnetic fields up to 68~T, we reach saturation of the anomalous Hall signal at a field $H_{\rm c} \simeq$ 55~T that was inaccessible in earlier studies, but is consistent with the expected N\'eel-vector reorientation field.

\end{abstract}

\maketitle

\section{Introduction}

The anomalous Hall effect (AHE) is a macroscopic linear response probe of time-reversal ($\cal{T}$) symmetry breaking in the electronic structure of magnetic materials \cite{Nagaosa2010,Smejkal2021b,Shindou2001,Machida2010,Chen2014,Kubler2014}. The established mechanisms include $\cal{T}$-symmetry breaking and AHE in conventional ferromagnets or in non-collinear magnetic structures \textcolor{black}{\cite{Nagaosa2010,Smejkal2021b,Shindou2001,Machida2010,Chen2014,Kubler2014,Nakatsuji2015,Kiyohara2016,Nayak2016}}. Recently, these have been extended by a novel and unconventional mechanism of $\cal{T}$-symmetry breaking resulting in prediction \cite{Smejkal2020,Smejkal2021b} and subsequently observation of the AHE \cite{Feng2022,Betancourt2021,Samanta2020} in a  class of  compensated collinear magnets whose opposite magnetic moments reside on crystal-sublattices connected by rotation symmetries \cite{Smejkal2021b,Smejkal2021a,Smejkal2022a}. Materials of this third unconventional class of collinear magnets, complementing the conventional ferromagnetic and antiferromagnetic classes, have a characteristic alternating spin polarization in both real-space crystal structure and momentum-space electronic structure that suggests the term altermagnets \cite{Smejkal2021a,Smejkal2022a}. The $\cal{T}$-symmetry breaking nature of the AHE thus implies that the sign of the AHE in altermagnets flips when reversing the sign of the N\'eel vector \cite{Smejkal2020,Smejkal2021b}.

In contrast to ferromagnets, where \textcolor{black}{AHE is always allowed}, the AHE in altermagnets can be allowed or excluded by symmetry depending on the \textcolor{black}{orientation} of the N\'eel vector with respect to the crystal axes \cite{Smejkal2020,Smejkal2021b}. \textcolor{black}{(For a comprehensive symmetry discussion of AHE in ferromagnets, altermagnets and non-collinear magnets we refer to recent review articles in Refs. \cite{Nagaosa2010, Smejkal2022a})} In altermagnetic MnTe, \textcolor{black}{for example, the AHE measured in the  (0001) c-plane of this hexagonal crystal is allowed for the N\'eel vector pointing along one of the equivalent \textcolor{black}{$\langle1\bar{1}00\rangle$} easy axes \cite{Feng2022,Betancourt2021}.} As a result, Hall measurements in the (0001) $c$-plane show a hysteretic AHE signal of opposite sign at opposite saturating fields with a coercivity on the order of a few T when sweeping the magnetic field along the $c$-axis, with a finite remanent AHE at zero field.

Altermagnetic RuO$_2$ shows a qualitatively distinct AHE phenomenology \cite{Smejkal2020,Feng2022}. For the N\'eel vector along the magnetic easy-axis, which corresponds to the [001] axis of this tetragonal rutile crystal, the AHE is excluded by symmetry \cite{Smejkal2020,Smejkal2021b}. An AHE is allowed by symmetry only when the N\'eel vector has a non-zero component in the (001) plane. As a result, no AHE signal is detected for a Hall bar patterned in the (001)-plane  when sweeping the out-of-plane field along the [001] $c$-axis \cite{Feng2022}. This sample and field geometry does not break the symmetry between opposite signs of the N\'eel-vector reorientation angle from the [001] easy axis \cite{Bazhan1976,Feng2022}. Therefore, even if the applied field was strong enough to cause a spin-flop reorientation of the N\'eel vector towards the (001)-plane, it would not generate a non-zero net AHE signal odd in the applied field \cite{Feng2022}. 

In contrast, a magnetic field applied along the [110] direction induces a continuous rotation of the N\'eel \textcolor{black}{vector by an angle $\alpha$ from the [001] toward the [110] axis}, given approximately by $\sin\alpha\sim H/H_{\rm c}$ \textcolor{black}{(for $ H < H_{\rm c} $)\cite{Bazhan1976,Feng2022}. Here, the direction of the N\'eel vector rotation is given by the sign of the magnetic field $H$, and $H_{\rm c}$ quantifies the field strength that aligns the N\'eel vector along the out-of-plane [110] axis.} The critical field $H_{\rm c}$ hereby depends on the exchange \textcolor{black}{interaction}, the magneto-crystalline anisotropy, and the Dzyaloshinskii-Moriya interaction (DMI) terms of the thermodynamic potential \cite{Bazhan1976} and in rutiles tends to be generally weaker than the spin-flop reorientation field applied along the [001] easy axis. In  RuO$_2$, \textcolor{black}{$H_{\rm c}$ } was estimated to be above 50~T, i.e. \textcolor{black}{beyond the field available in the earlier AHE study \cite{Feng2022}, but} below the [001] spin-flop field whose strength was estimated to exceed 100~T \cite{Feng2022}.

In the recent study \cite{Feng2022}, a non-linear Hall signal at high fields in \ch{RuO2} Hall bars patterned in the (110)-plane was interpreted as a strong, yet unsaturated, AHE generated by the reorientation of the  N\'eel vector into the (1$\bar{1}$0)-plane by the applied [110] field. The opposite \textcolor{black}{direction} of the N\'eel vector rotation for opposite applied fields gave an opposite sign of the AHE~\cite{Feng2022}. The detected AHE signal was, therefore, odd in the applied [110]-field, and vanished at zero field. A comparison to the measurements in the (001)-plane Hall bars evidenced a strong AHE contribution in the (110)-plane samples, dominating the ordinary Hall contribution over the whole field range up to 50~T~\cite{Feng2022}. However, as $H_{\rm c}$ exceeded the experimentally available magnetic field range, no saturation of the AHE has been observed. \textcolor{black}{The appearance of a saturation would make a strong case for an AHE generated by the N\'eel vector reorientation, for which such a saturation above $H_{\rm c}$ is expected.}

In this study, we present magnetotransport measurements on (110)-plane oriented RuO$_2$ samples analogous to those explored in Ref.~\cite{Feng2022}. \textcolor{black}{We further optimized the film preparation protocol to achieve high crystalline quality and stoichiometry, as evidenced by a thorough structural characterization. We performed on these samples a systematic magnetic and magneto-transport characterization up to magnetic fields of 68~T. Our results demonstrate consistency with Ref.~\cite{Feng2022} below 50~T, and by going beyond 50~T magnetic fields allows us to reach the saturation of the AHE signal. The observation of a saturation completes the experimental evidence of the AHE in this workhorse altermagnet \cite{Smejkal2020,Feng2022,Bose2022,Bai2021,Karube2022}.}

\section{Thin film growth and characterisation}

Our RuO$_2$(110) samples were grown epitaxially on (110)-oriented substrates of rutile TiO$_2$ (CRYSTAL GmbH) using a commercial pulsed laser deposition (PLD) setup (TSST B.V.) with a pulsed excimer laser (COMPex Pro 205/KrF, 248~nm). Prior to growth, the substrates were cleaned in subsequent ultrasonic baths of isopropanol and acetone (20 min, puriss.) and annealed for 5~h in a tube furnace at 820~$^{\circ}$C with an oxygen flow of 20~L/h to obtain a stepped-terrace morphology. The PLD growth conditions were set to an oxygen partial pressure of $1\times 10^{-3}$~mbar and a substrate temperature of 700~K (two-color pyrometer, IMPAC). Then, the \textcolor{black}{sintered} RuO$_2$ powder target (TOSHIMA manufacturing Co. Ltd.) was ablated by 32.000 laser pulses at a repetition rate of 10 Hz. The laser energy density was gradually increased from 0.7 to 1.4~J/cm$^2$ to compensate for \textcolor{black}{laser induced} oxygen deficiencies and a consequent increase of target reflectivity during deposition. 
The film growth was monitored by reflective high-energy electron diffraction (RHEED, STAIB) and post characterized by X-ray reflectivity (XRR) \textcolor{black}{and X-ray diffraction (XRD) on a Rigaku Smartlab rotating anode with parabolic mirror and Ge channel cut monochromator using} 8047.8~eV Cu K$\alpha$ radiation, \textcolor{black}{as well as} X-ray photoelectron spectroscopy (XPS, OMICRON, 1486.6~eV Al K$\alpha$; \textcolor{black}{data not shown) to check the stoichiometry}. 

The \textcolor{black}{film characterization} results are  shown in Fig.~1. \textcolor{black}{Fig.~1a shows an XRD close-up around the 110 Bragg peak of the \ch{TiO2} substrate and the epitaxial \ch{RuO2} thin film, taken from the full range scan in the inset, and providing lattice parameters that are in excellent agreement with the literature values ~\cite{Feng2022}.}
\textcolor{black}{The data are further compared to} simulations \textcolor{black}{for a 9.8~nm thick film with 0.5~nm roughness, obtained via} the Parrat formalism and a dynamic diffraction model as implemented in Ref.~\cite{Kriegner:rg5038}.

The \textcolor{black}{XRR} in Fig.~1b confirms the thickness of our sample \textcolor{black}{to be} $\approx9-10$~nm. \textcolor{black}{Both XRR and XRD show thickness fringes corresponding to the same thickness, proving excellent crystalline quality with a coherent crystal lattice from the interface with the substrate to the surface and correspondingly low roughness.} 
The crystalline quality is further confirmed by transmission electron microscopy (TEM, FEI Titan 80-300, U = 80-300~kV, I $>$ 0.6~nA) in Fig.~1d.

The low roughness \textcolor{black}{suggested by XRD and XRR} is consistently demonstrated by scanning tunneling microscopy (STM) images \textcolor{black}{(Omicron VT-STM, constant current mode, RT, U = 500~mV, I = 0.05~nA)} in Fig.~1c, yielding a rms of $<$~2~nm. 

\textcolor{black}{A superconducting quantum interference device (SQUID) MPMS3} magnetometry measurement of our sample at 5~K is shown in Fig.~1e. A measurement on a bare TiO$_2$ substrate is used \textcolor{black}{as a reference} to subtract the background from the data measured on the 9~nm RuO$_2$ film. \textcolor{black}{As the signal from the magnetic \ch{RuO2} moments is extremely small as compared to the \ch{TiO2} background}, the presented magnetometry data \textcolor{black}{is subject to a large} error bar. \textcolor{black}{However,} the weak field-induced moment is consistent with the previous report by Feng et al.~\cite{Feng2022}.

\section{Magneto-transport}

To perform transport measurements, Hall bars were fabricated by e-beam lithography out of the \ch{RuO2} thin films. Trenches defining the Hall bars were etched by argon ion beam etching. The Hall bars have different orientations in the sample (110) plane, either along the [001] or the [1$\bar{1}$1] axes, as shown in the inset of Fig.~2a. All Hall bars have a width of 10~$\mu$m, with distances between contacts of 50 and 100~$\mu$m, as shown in the second inset of Fig.~2a. 

The magneto-transport measurements were first performed in static magnetic field in a standard \textcolor{black}{Quantum Design Physical Properties Measurement System} (QD PPMS) setup and afterwards the same device was studied in the high magnetic field laboratory \textcolor{black}{(HLD) in Dresden-Rossendorf} using pulsed magnetic fields up to 68~T, \textcolor{black}{with pulses of 100~ms duration and a rise time of 35~ms}. The magnetic field was applied in the out-of-plane sample direction, which corresponds to the RuO$_2$ [110] crystalline direction. \textcolor{black}{To investigate the angular dependence of the Hall resistivity, the magnetic field was additionally tilted by $\theta=\pm$45$^{\circ}$ towards [001], as illustrated in the inset of Fig. 4.}. An \textcolor{black}{alternating current} (AC) with an amplitude of 5~$\mu$A was applied and the voltage was measured using a numerical lock-in technique \textcolor{black}{with frequencies in the kHz range} to suppress the noise level and spurious effects from the pulsed magnetic field. The presented high magnetic field data were measured on two samples grown under the same conditions showing reproducible results.

\textcolor{black}{Fig.~2a shows the temperature dependent longitudinal resistivity $\rho$, which decreases with decreasing temperature, indicating metallic behavior in the thin films. 
The main result of our study is the Hall resistivity  $\rho_{\rm Hall}$ measured up to 68~T, which is shown in Fig.~2b. The measurement in pulsed magnetic fields results in higher noise levels, therefore the curve was smoothed using a Savitzky-Golay filter and the noise is indicated by the error-bar. The Hall resistivity shows a pronounced non-linearity, exhibiting a non-zero curvature beyond 20~T. This is consistent with the measurements on \ch{RuO2}(110) films reported in Ref. ~\cite{Feng2022}. However, with the maximum accessible magnetic field of 50~T, a clear signature of the saturation of the AHE signal was not observed in Ref. ~\cite{Feng2022}.
In our data in Fig.~2b, we observe the saturation of the AHE above $H_{\rm c} \simeq$ 55~T. This field range is consistent with the expected 50~T-100~T scale of the reorientation field of the N\'eel vector from the [001] easy axis to the [110] hard axis \cite{Feng2022}.
Moreover, the very weak asymptotic slope above $H_{\rm c}$ (corresponding to the ordinary Hall effect), together with the overall magnitude of $\rho_{\rm Hall}$, shows that the AHE signal dominates over the ordinary Hall contribution over the entire range of applied magnetic fields.}

\textcolor{black}{Temperature-dependent magneto-transport data are shown in Fig.~3. The longitudinal magnetoresistance in Fig.~3a reaches 15\% at 68~T and low temperature. At low fields, the magnetoresistance is roughly parabolic, and the amplitude of the magnetoresistance strongly decreases with increasing temperature. The temperature-dependent Hall resistivity is shown in Fig.~3b. The overall $\rho_{\rm Hall}$ amplitude decreases with increasing temperature, up to our maximum temperature of 200~K. In addition, we see a clear saturation of the Hall signal at temperatures up to 80~K. }

\textcolor{black}{Fig.~3c highlights the decrease in the Hall resistivity at 68 T with increasing temperature for two Hall bars patterned along the [001] and [1$\bar{1}$1] directions. The data of both samples are in remarkable agreement, indicating the N\'eel vector in both cases to be rotated into the (1$\bar{1}$0) plane upon application of the [110] oriented magnetic field, as illustrated in the inset of Fig. 2a.}

\textcolor{black}{To further investigate the influence of a magnetic field component applied along the [001] easy axis on the AHE, we measure the Hall resistivity while tilting the field away from the [110] direction toward the [001] direction. The results are presented in Fig.~4, showing the Hall resistivity as a function of the [110] field component $H_{\perp}$ for tilt angles of $\theta=\pm$45$^\circ$. \textcolor{black}{Again, all datasets are in remarkable agreement confirming that} the AHE solely depends on the projection of the field onto the [110] axis. \textcolor{black}{Most importantly}, this implies that applying a magnetic field along the easy [001] axis does not tilt the N\'eel vector into the (110) plane, as this would result in saturation at lower fields.}

\section{Conclusion}

In conclusion, we have systematically investigated structural, magnetic and magneto-transport characteristics of epitaxial thin-film \ch{RuO$_2$}(110) grown on \ch{TiO$_2$}(110) substrates. The excellent crystalline quality of our films is demonstrated by XRR, TEM and STM measurements. Our magnetotransport results below 50~T are qualitatively consistent with the earlier report in Ref.~\cite{Feng2022} \textcolor{black}{and provide an additional confirmation of the unconventional AHE originating from collinearly compensated magnetic order in altermagnetic RuO$_2$.}

\textcolor{black}{Going beyond a mere confirmation, our work augments the range of accessible magnetic fields up to 68 T, fields that are large enough to saturate the AHE resistivity. The corresponding saturation field is consistent with earlier estimates~\cite{Feng2022}.}

\textbf{Acknowledgements} 
\textcolor{black}{L.V. was supported by the Leibniz Association through the Leibniz Competition. Further funding support came from the Deutsche Forschungsgemeinschaft (DFG, German Research Foundation) under Germany's Excellence Strategy through the W\"urzburg-Dresden Cluster of Excellence on Complexity and Topology in Quantum Matter ct.qmat (EXC 2147, Project ID 390858490) as well as through the Collaborative Research Center SFB 1170 ToCoTronics (Project ID 258499086). We further acknowledge funding support of the Czech Science Foundation Grant Nos. 19-18623X, GACR 22-17899K and DFG 490730630. Finally, this work was supported by HLD-HZDR, a member of the European Magnetic Field Laboratory (EMFL).}


%

\begin{figure*}
	\includegraphics[width=1\textwidth]{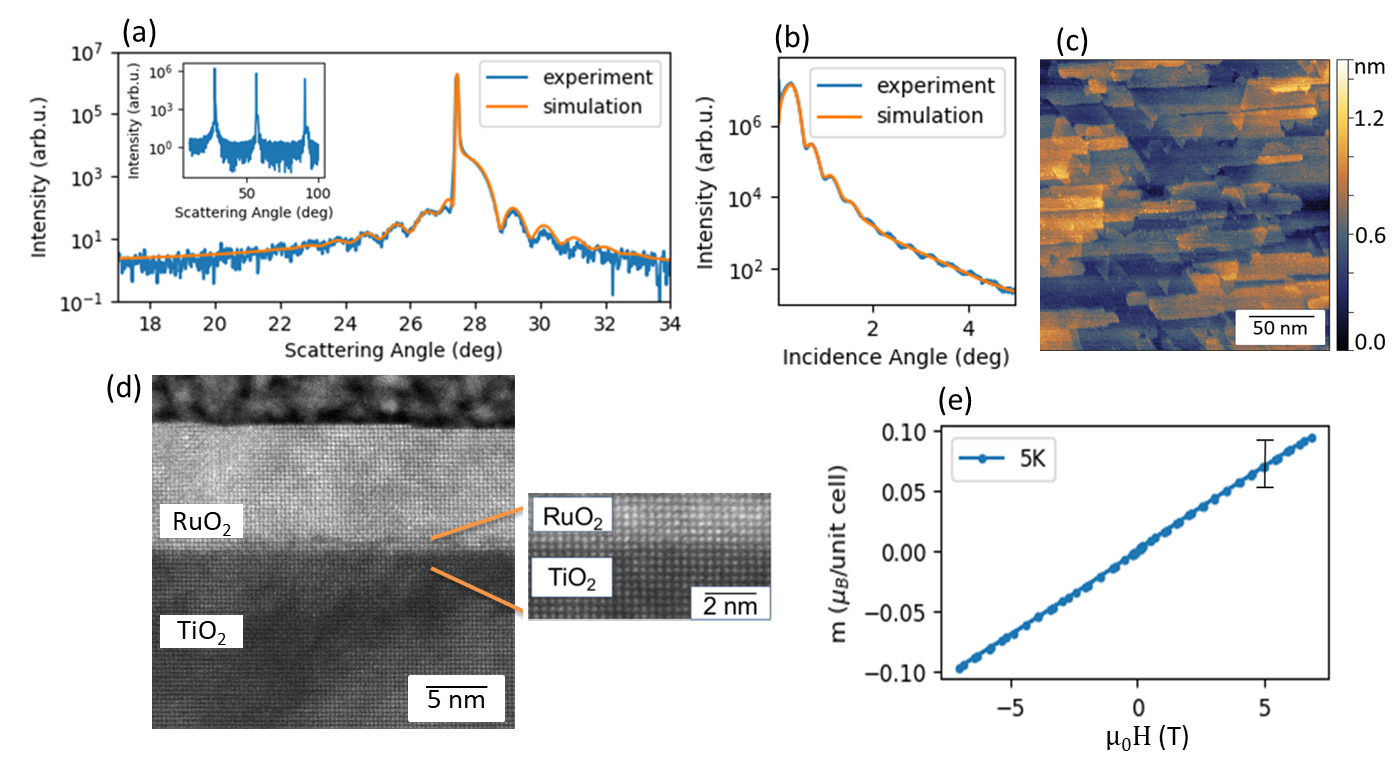}
	\caption{\textbf{Growth and characterisation of RuO$_2$ thin film on a TiO$_2$(110)}.(a) \textcolor{black}{XRD data of the 110 Bragg peak} \textcolor{black}{(full range in the inset) and (b) \textcolor{black}{XRR} show good agreement with} simulations for a 9.8~nm thick film with 0.5~nm roughness. (c) \textcolor{black}{STM} shows a smooth film surface. (d) \textcolor{black}{TEM} shows \textcolor{black}{epitaxial and dislocation-free growth of RuO$_2$(110) on TiO$_2$(110)}. (e) \textcolor{black}{SQUID} magnetometery measurements  on the RuO$_2$/TiO$_2$ sample with subtracted background based on a TiO$_2$ reference measurement. \textcolor{black}{The large error bar of the magnetization data is \textcolor{black}{due to the weak field-induced magnetic moments in RuO$_2$}}}
	\label{Figure1}
\end{figure*}


\begin{figure*}
	\includegraphics[width=0.95\textwidth]{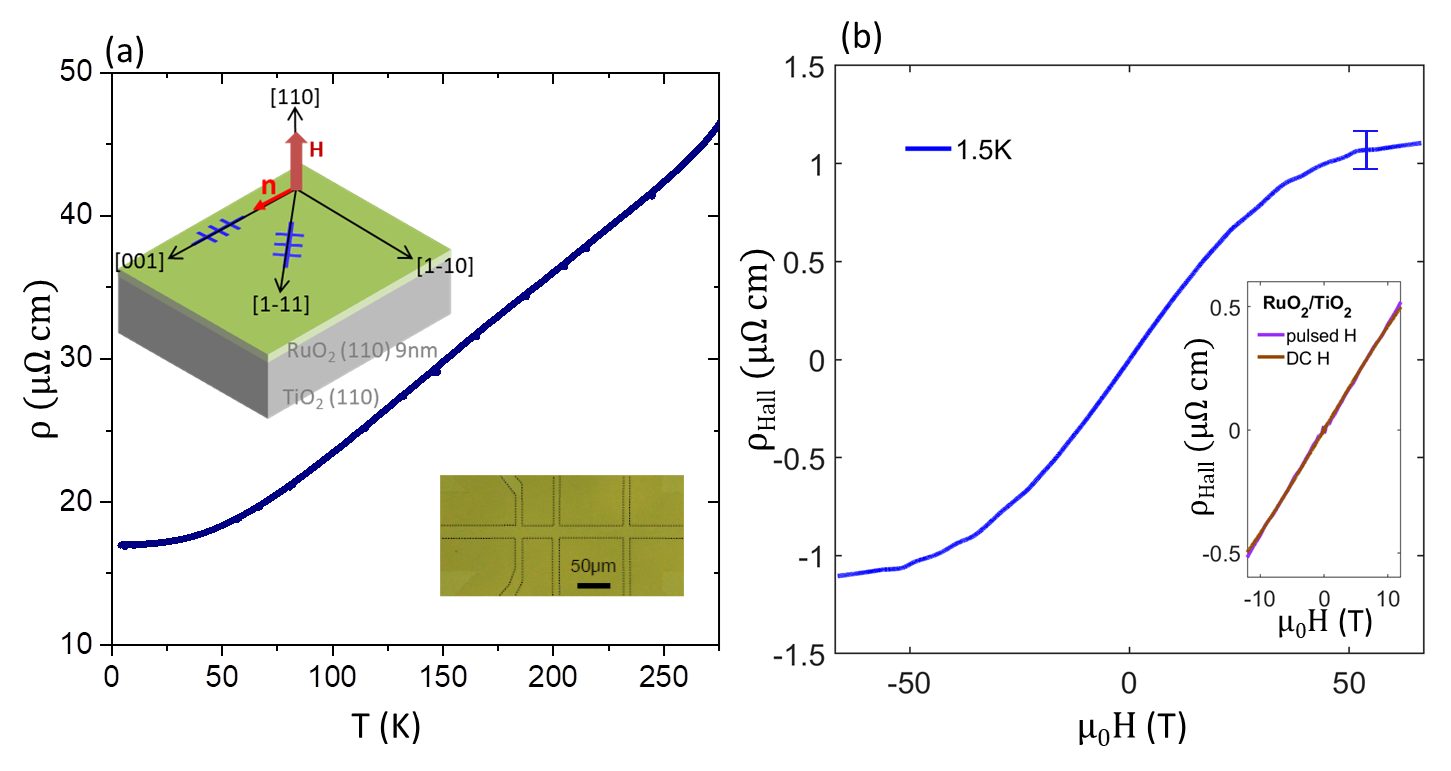}
	\caption{\textbf{\textcolor{black}{Magneto-transport in} (110)-oriented \ch{RuO2} \textcolor{black}{thin films}}. (a) Temperature dependence of the longitudinal resistivity showing metallic behavior. Upper inset: Orientation of the Hall bars with respect to the crystal directions and the directions of the N\'eel-vector easy axis ({\bf n}) and applied field axis ({\bf H}). Lower inset: optical microscopy image of the \ch{RuO2} Hall bar, highlighted by dashed lines. (b) Antisymmetrized transverse resistivity (Hall resistivity) of a \ch{RuO2} Hall bar oriented in the [001] direction \textcolor{black}{measured} in the high magnetic field facility up to 68~T. \textcolor{black}{The inset compares the pulsed field to the in house static field measurements up to 12~T.}}
	\label{Figure2}
\end{figure*}

\begin{figure*}
\centering
	\includegraphics[width=1\linewidth]{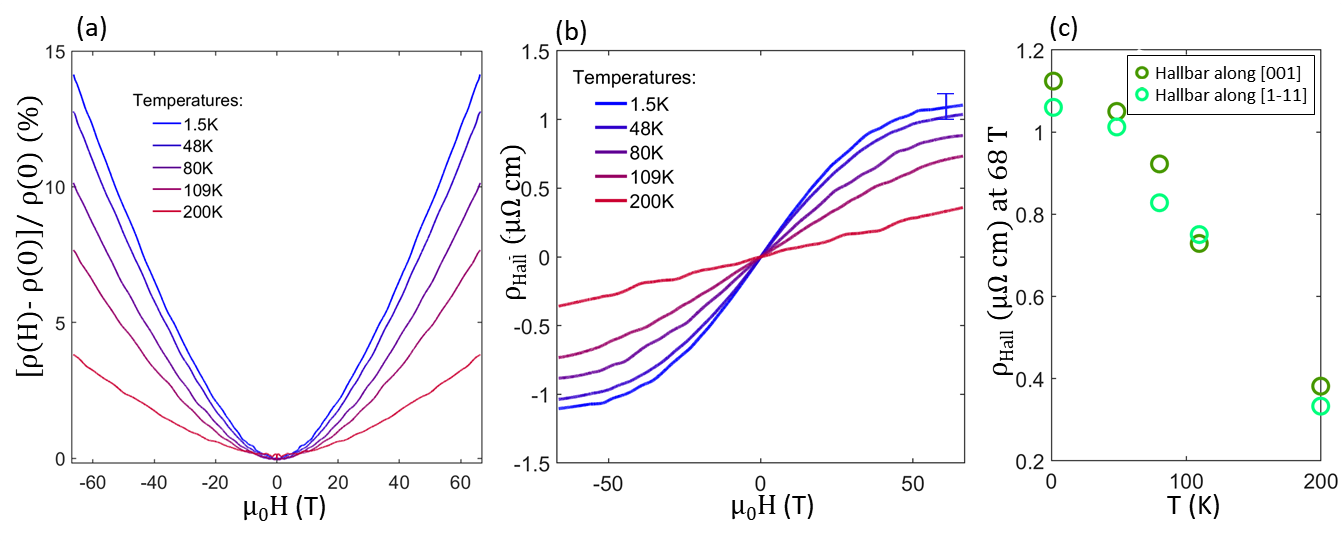}
	\caption{\textbf{Temperature dependent magneto-transport data \textcolor{black}{of RuO$_2$(110)}.} (a) \textcolor{black} {Temperature dependence of the} symmetrized longitudinal magnetoresistance and (b) the antisymmetrized transverse resistivity (Hall resistivity). \textcolor{black}{The error-bar indicates the uncertainties caused by noise in high magnetic fields.} \textcolor{black}{(c) Comparison of the temperature dependencies of the Hall resistivity at B=68 T for Hall bars oriented along the [001] and [1$\bar{1}$1] directions, respectively}. }
	\label{Figure3}
\end{figure*}

\begin{figure*}
\centering
	\includegraphics[width=0.5\linewidth]{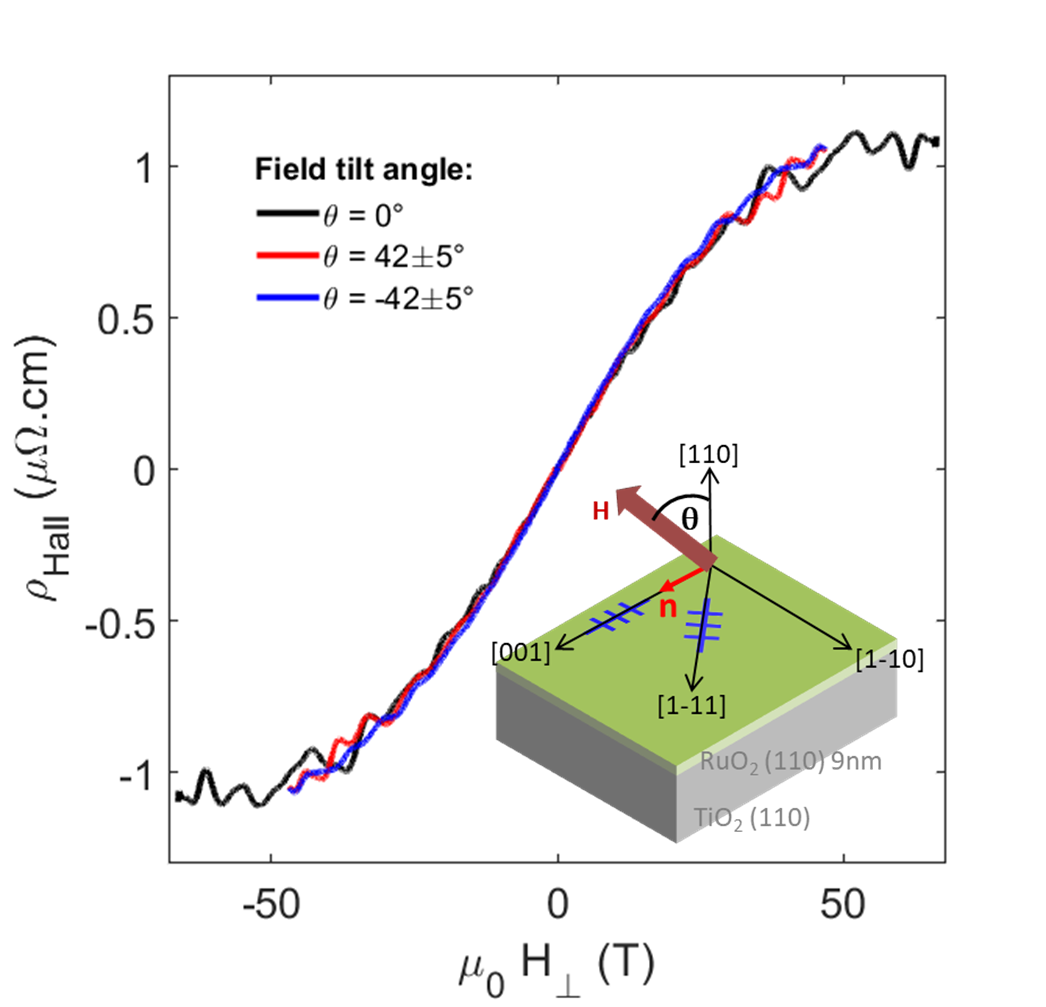}
	\caption{\textcolor{black}{\textbf{Effect of magnetic field tilting on the Hall resistivity in RuO$_2$(110).} The Hall resistivity is shown with respect to the out-of-plane component of the tilted magnetic field. The magnetic field was rotated by $\pm$45$^{\circ}$ toward the [001] axis. The inset shows the orientation of the Hall bars with respect to the crystal directions of the N\'eel-vector easy axis ({\bf n}) and the applied field axis ({\bf H}) rotated at an angle $\theta$.}}
	\label{Figure4}
\end{figure*}

\end{document}